\documentstyle[aps,prb,twocolumn]{revtex}
\begin{document}

\begin{center}
{\Large Interaction-induced chaos in a two-electron quantum-dot system}

\bigskip
 A.\ J.\ Fendrik, M.\ J.\ S\'anchez and  P.\ I.\ Tamborenea \\
Departamento de F\'{\i}sica J.\ J.\ Giambiagi\\
Facultad de Ciencias Exactas y Naturales \\
Universidad de Buenos Aires \\
(1428) Buenos Aires, Argentina

\end{center}

\bigskip\noindent
{\it 
A quasi-one-dimensional quantum dot containing two interacting electrons 
is analyzed in search of signatures of chaos.
The two-electron energy spectrum is obtained by diagonalization
of the Hamiltonian including the exact Coulomb interaction.
We find that the level-spacing fluctuations follow closely a Wigner-Dyson
distribution, which indicates the emergence of quantum signatures of chaos
due to the Coulomb interaction in an otherwise non-chaotic system.
In general, the Poincar\'e maps of a classical analog of this quantum 
mechanical problem can exhibit  a  mixed classical dynamics.
However, for  the range of  energies involved in the present system, 
the dynamics is strongly chaotic, aside from  small regular regions.
The system we study models a realistic semiconductor nanostructure,
with electronic parameters typical of gallium arsenide.
}


\bigskip\noindent

In a many-body system, it is possible for signatures of quantum chaos to 
appear due solely to the interactions among its particles.
During the last decade, such interaction-induced signatures of quantum chaos
have been investigated, for example, in spin-fermion 
models,\cite{mon-poi-bel-sir} in the compound nuclear state with 12 
particles in 
the  $sd$-shell\cite{zel-hor-bro} and in the heavy rare-earth atom of 
Cerium.\cite{fla-gri-gri-koz}
In those  studies, the evidence for quantum signatures of chaos found 
in the level-spacing statistics or in the statistical properties of the wave 
functions was considered to be conclusive.
This conclusion is not entirely surprising since all of those systems,
having relatively large numbers of particles ($>10$), are similar to the 
complex nuclear systems that motivated the introduction of the ideas of
random matrix theory (RMT) in the first place.\cite{dyson}
On the other hand it is not obvious a priori, whether the inclusion
of the interaction in  few-body systems, like for example currently 
available semiconductor quantum-dots, leads also to  signatures
of quantum chaos. 
In a recent study, the effect of electronic interactions was considered in 
a parabolically-confined three-electron system.\cite{mes-ull-pfa}
(It is well known that three is the lowest number of interacting electrons 
necessary to break the integrability  in a parabolic quantum dot.) 
In that study, the crossover from regular to irregular spectra as a function 
of  the interaction strength was found to be incomplete, possibly due to 
the existence of hidden symmetries not taken into account in the 
statistical analysis.
Therefore, the question of the emergence of signatures of quantum chaos 
due to interparticle interactions in systems of very few particles remains 
open. 

Closely related to the issue of characterizing the dynamical properties of
simple interacting systems is the problem of quantum control with external 
fields.
The manipulation of few particles (electrons and holes) in semiconductor 
quantum dots is a potentially important technological problem that 
is receiving increasing attention.\cite{lun-sch-lee-pet,luyken-etal}
In this context, recent theoretical studies have shown interesting effects of
single-electron turnstile behavior, and localization and correlations
in systems of quantum dots with two interacting electrons in 
them.\cite{tam-met}
In the present Letter we investigate the signatures of quantum chaos in
a similar system, ie., two interacting electrons in a quasi-one-dimensional
semiconductor quantum dot.
We show that the Coulomb interaction between the electrons induces an 
unambiguos transition from a  regular spectrum  to
a spectrum  that follows closely the predictions of RMT for systems 
whose classical analog exhibits chaotic dynamics.

In order to fully characterize the emergence of chaos due to interactions
in this simple system, we also study the dynamics of its classical analog.
The Poincar\'e maps that describe the classical system  show a 
strongly (albeit mixed) chaotic behavior due to the inclusion 
of the true Coulomb interaction in the system.

We assume that the quantum dot has narrow parabolic confinement in
the transversal $x-y$ dimensions, so that the energies associated 
with those modes  are high compared to the energies of the 
remaining degree of freedom (Born approximation).
The two-electron wave function can then be written as
\begin{equation}
\Psi({\mathbf r}_1,{\mathbf r}_2)=\phi(x_1) \phi(y_1) \phi(x_2) 
\phi(y_2) \Phi(z_1,z_2),
\end{equation}
where $\phi(x)$ is the lowest harmonic oscillator energy eigenstate.
The energy eigenstates satisfy
\begin{eqnarray}
[ H_0(z_1) + H_0(z_2) + V_{1D}(|z_1-z_2|)] \Phi(z_1,z_2) \nonumber \\
= E \Phi(z_1,z_2),
\label{eq:sch}
\end{eqnarray}
where
$H_0(z)=-(\hbar^2/2m^{\ast}) \; \partial^2/\partial z^2 + V(z)$
is the single-particle Hamiltonian with $V(z)$ being the quantum-dot 
defining potential. 
$m^{\ast}$ is the effective mass, and $V_{1D}$ is the Coulomb 
interaction given by
\begin{eqnarray}
V_{1D}(|z_1-z_2|) &=& \int dx_1 dy_1 dx_2 dy_2 \nonumber \\
& & \frac{e^2 \phi^2(x_1) \phi^2(y_1) \phi^2(x_2) \phi^2(y_2)}
     {\epsilon |{\mathbf r}_1-{\mathbf r}_2|}.
\label{eq:V1d}
\end{eqnarray}
We use the values of the dielectric constant $\epsilon=13.1$ and 
$m^{\ast}=0.067 m_e$ corresponding to gallium arsenide.
We choose to work with a quasi-one-dimensional semiconductor quantum dot 
confined  in  $15~\AA$ in the x-y plane and a width of
$800~\AA$ in the z-direction.

In the absence of the interaction term in Eq.~(\ref{eq:sch}), the
Hamiltonian is a sum of two single-particle one-dimensional 
Hamiltonians, whose classical counterpart is obviously integrable. 
The main question we seek to answer is whether the 
Coulomb interaction between the electrons introduces chaos in the system.

In order to look for signatures of quantum chaos, we follow a standard
statistical analysis of the energy spectrum, which consists of the
following steps.
First we calculate the exact spectrum $\{E_n\}$ by diagonalization of the
Hamiltonian matrix.   
The level spectrum is used to obtain the smoothed counting function 
$N_{av}(E)$ which gives the cumulative number of states below an 
energy $E$. 
In order to analize the structure of the level fluctuations 
properties one ``unfolds" the spectrum by applying the well 
kwown transformation $x_n=N_{av}(E_n)$.\cite{boh}
From the unfolded spectrum one calculates the nearest-neighbor spacing (NNS)
distribution $P(s)$, where $s_i \equiv x_{i+1}-x_i$ is the NNS.

We first consider the spectral properties of the non-interacting two-electron
problem whose Hamiltonian is $H_0(z_1)+H_0(z_2)$. 
Its eigenstates can be classified by their total spin in singlets and 
triplets.
To compute the NNS distribution we use eigenstates of a given spin.
In the inset of Fig.~1 we show the obtained NNS non-interacting distribution 
$P_{NI}(s)$ (histogram) which follows a Poisson distribution (characteristic
of  an  uncorrelated sequence of energy levels) given by 
$P_P(s)=e^{-s}$ and shown for comparison as a solid thin line.
Due to the finite dimension of the Hilbert space, $N_{av}(E)$ saturates 
in the highest energy region. 
Therefore, we compute the NNS distribution using the lowest $\sim 1000$ 
eigenvalues.\cite{note}
The obtained Poisson distribution is an expected signature of most quantum 
two-dimensional systems whose classical counterparts are 
integrable.\cite{ber-tab}

To analize the interacting spectrum we diagonalize exactly Eq.~(\ref{eq:sch}). 
We also take into account the symmetry of the spectrum due to the parity 
of the confining potential and the interaction potential.
Therefore, in order to compute the NNS distribution we use eigenstates of 
a given parity and spin.
This kind of decomposition is a standard procedure followed in the 
analysis of spectral properties of quantum systems whenever the 
Hamiltonian of the system possesses a discrete symmetry.\cite{boh}
After unfolding the spectrum the NNS distribution is computed for the even 
parity states. 
Again, we consider $\sim 1000$ eigenstates of the interacting Hamiltonian 
(whose eigenenergies are lower than the energy of the first transversal 
mode, for compatibility with the Born approximation). 

Since the singlet is the ground state of the two-electron system, we first 
concentrate on the subspace of spatial wave functions that are symmetric 
under particle exchange.
The interaction affects very clearly the spectrum, resulting in a strong 
level repulsion: the NNS distribution is in accordance with the predictions
of RMT.\cite{mehta} 
As a consequence, the obtained NNS distribution $P_{IS} (s)$ (histogram 
shown as a thick solid line in Fig.~1) is well described by the Wigner 
surmise\cite{mehta}
$P_W (s)=\frac{1}{2} \pi \,s  \exp ({- \pi s^2 /4})$,
shown for comparison as a thin solid line.

For the triplet states, due to the antisymmetry of the spatial wave functions,
it is reasonable to expect that the tendency of the two electrons to avoid 
each other results in a weaker level mixing.
Nevertheless, although some differences with the singlet case would appear 
for other statistical measures (that are not possible to perform with 
the number of levels at hand), those differences are  not qualitatively 
visible on the computed NNS distribution $P_{IT} (s)$, shown as a dashed line
in Fig.~1. Again the obtained  histogram fits quite well with the predictions 
of RMT.

\vspace{.5cm}
\begin{figure}
 \vbox to 8cm {\vss\hbox to 6cm
 {\hss\
   {\includegraphics{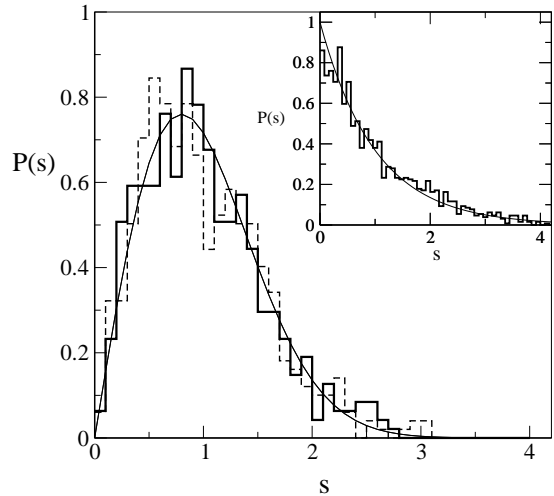}
   }
  \hss}
 }
\caption{NNS distribution obtained for the interacting two-electron system 
described in the text. The solid thick line (dashed line) is $P_{IS} (s)$ 
$(P_{IT} (s))$. The  Wigner surmise $P_{W} (s)$ is plotted as a thin solid 
line in  order to compare it  to the obtained distributions.
Inset: NNS distribution $P_{NI} (s)$ for the 
non-interacting two-electron systems described in the text (thick line) 
togheter with the Poisson distribution $P_P (s)$ (thin line).}
\end{figure}

We now turn to the dynamics of the classical counterpart of the two-electron
quantum dot system.
We consider the classical interaction potential given by
$V_{cl}(|z_1 - z_2|) = \alpha / \sqrt{( d^2 +  |z_1-z_2|^2 )}$, 
where the parameters $\alpha$ and $d$ have been obtained from the best 
fit to the Coulomb interaction $V_{1D}(|z_1-z_2|)$, Eq.~(\ref{eq:V1d}).
(The fit is very good at all distances, down to the resolution of the spatial 
grid used in our numerical calculations.)
The classical single particle confining potential $V_{C}(z)$ is a square well 
of length $L$, and we restrict the analysis to bounded motion within this
box.
In this situation, the effect of the confining potential is to reflect 
elastically the particles off the boundaries in each bounce, breaking the 
translational symmetry of the problem. 
As a consequence, the center of mass momentum is not preserved.
Nevertheless, in the absence of interaction each  single particle energy  
is a constant of motion, and therefore the classical problem 
is integrable.
On the other hand, the inclusion of the Coulomb interaction 
breaks the conservation of the
single particle  energies and can induce  an irregular dymamics in the
confined system.

For a total energy $E = E^{*} \equiv \alpha / d $ there is a separatrix 
in the classical dynamics. 
That is, for $E < E^{*}$ the particles never cross each other and the sign 
of the relative coordinate $z_2 -z_1$ never changes, while for $E > E^{*}$ 
it can change. 

Denoting $\epsilon \equiv E / E^{*}$ we write 
\begin{equation}
\label{ch}
\epsilon = \frac{{v_{1}^{\prime}}^{2}}{2} + \; \frac{{v_{2}^{\prime}}^{2}}{2} +\; V_C (z_1^ {\prime}) + \; V_C (z_2^{ \prime}) + \frac{d^*}{\sqrt{{d^*}^{2} + 
| z_1 ^{\prime} -  z_2 ^{\prime}|^2}} \; ,
\end{equation}
where we have defined 
\begin{equation}
\label{const}
v_i^{\prime} =  v_i  \frac{\sqrt{ m^* d}}{\sqrt\alpha} \;, \; \; 
z_i^{\prime}  =  \frac{z_i}{L}   \; ,\;\;  
d^{*} =   \frac{d}{L} \; , 
\end{equation}
with $i = 1,2$. 
In this way, for a given value of the reescaled energy $\epsilon$, the 
classical dynamics depends only on the parameter $d^{\ast}$. 
Taking into account that $L= 800 \AA$ and the best fit with the quantum 
Coulomb term $V_{1D}$ gives $d= 8 \AA$, we obtain $d^{\ast}= 0.01$.

In Fig.~2(a) we show for $\epsilon=0.9$ the Poincar\'e surface of section 
$v_{2} ^{\prime}$ vs. $z_{2}^{\prime}$, for the motion of
one of the particles, taken at times when the other particle bounces off 
the left boundary of the well (the topology of the Poincar\'e section does not 
depend on which particle is selected). 
The motion is chaotic over most of the accesible phase space for the given 
energy shell.
Fig.~3(a) shows another Poincar\'e section for  $\epsilon= 16$.
Again, except for small regions of regular motion, the dynamic is 
fully chaotic. 

Although the two  Poincar\'e sections look qualitatively similar, 
the trajectories in the plane  $z_{2} ^{\prime}$ vs. $z_{1}^{\prime}$ are 
quite different as can be seen from Figs.~2(b) and 3(b).
Figure 2(b) (3(b)) shows for $\epsilon=0.9$ ($\epsilon=16$) a piece of a
trajectory in the  $z_{2} ^{\prime}$ vs. $z_{1}^{\prime}$ plane 
corresponding to an initial condition in the chaotic region.
In Fig.~2(b) the trajectory never crosses the straight line defined by 
$z_{2} ^{\prime} = z_{1}^{\prime}$, because for  $\epsilon=0.9$ is always
$z_{2}^{\prime} > z_{1}^{\prime}$. 
In Ref.\onlinecite{meza} the authors perform a classical 
analysis of  the emergence of chaos due to the inclusion of an interparticle
screened Coulomb interaction in an infinite well. 
The classical motion  of such a system is  qualitatively similar 
to  our classical model only for $\epsilon  < 1$.

From the $\approx 1000$  eigenstates employed to compute the NNS distribution 
displayed in  Fig.~1, only  the lowest 5~\% have eigenenergies
that correspond to   $\epsilon  < 1$.
The eigenenergies of the remaining states correspond to  values of 
$\epsilon$ ranging from 1 to 16, for which, as we have shown, the classical
dynamics is chaotic over most part of the energy shell. 
Therefore the NNS distribution computed from these 
states results in  a remarkable quantum signature of the  
underlying classical chaotic dynamics.
 
For energies $\epsilon  >> 1$ the classical regular regions in the Poincar\'e 
maps should become predominant over the chaotic ones, because in such a limit  
the Coulomb term can be considered as  a perturbation of  the 
non-interacting two-electron Hamiltonian. 
Nevertheless, values of  
$\epsilon  >> 1$ are not realistic for quantum wells 
describing semiconductors nanoestructures.

\begin{figure}
 \vbox to 8cm {\vss\hbox to 6cm
 {\hss\
   {\includegraphics{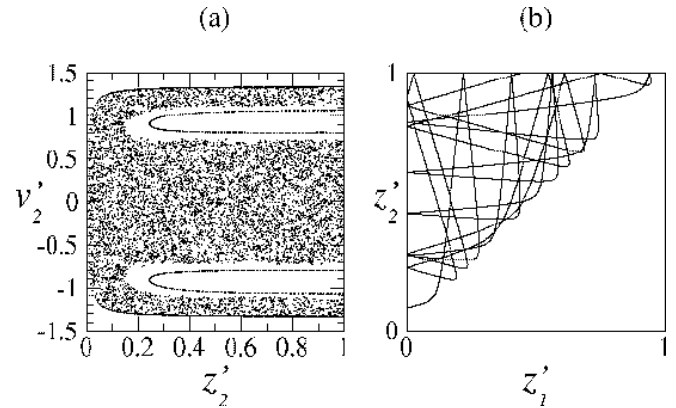}
   }
  \hss}
 }
\caption{(a) Poincar\'e surface of section 
$v_{2} ^{\prime}$ vs. $z_{2}^{\prime}$ for a rescaled energy $\epsilon = 0.9$.
(b) For the same value of $\epsilon$, a piece of a  trajectory in the 
$z_{2} ^{\prime}$ vs. $z_{1}^{\prime}$ plane corresponding to an initial 
condition in the chaotic region.}
\end{figure}

\begin{figure}
 \vbox to 8cm {\vss\hbox to 6cm
 {\hss\
   {\includegraphics{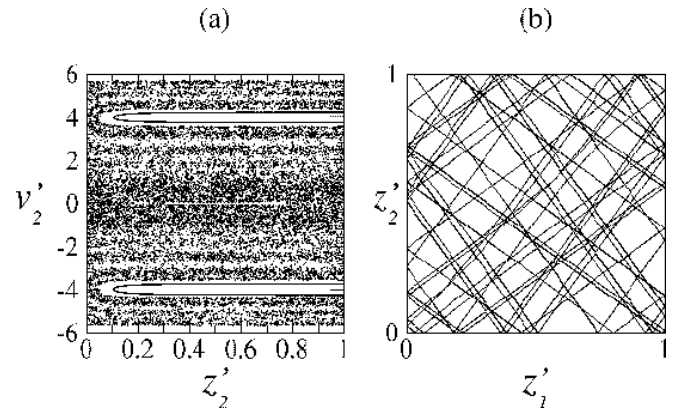}
   }
  \hss}
 }
\caption{(a) Poincar\'e surface of section 
$v_{2} ^{\prime}$ vs. $z_{2}^{\prime}$  for a rescaled energy $\epsilon = 16$.
(b) For the same value of $\epsilon$, a piece of a  trajectory in the 
$z_{2} ^{\prime}$ vs. $z_{1}^{\prime}$ plane corresponding to an initial 
condition in the chaotic region.}
\end{figure}

In conclusion, we show that the  Coulomb interaction is responsible
for the chaotic dynamics in  a quasi-one-dimensional  two-electron-quantum 
dot. This is the first study of a realistic
few body system where the emergence of chaos due to interparticle interactions
is unmistably demostrated through 
the analyses  of both its quantum  spectral properties and the dynamics of its
classical counterpart.
We believe that the present results also put serious constraints to 
some models of semiconductors nanoestructures in which the interaction 
among particles is modeled, for a finite number of particles, as a capacitive 
term  in the form of a constant interaction (CI).
The inclusion of the CI  gives an  interacting 
Hamiltonian whose  spectral properties are those of the non-interacting one.
In other words, a Poisson  NNS distribution will remain Poissonian after 
considering  the interaction in the CI model.

This work was partially supported by UBACYT (TW35), PICT97 03-00050-01015,
Fundaci\'on Antorchas and CONICET.


\end{document}